# Ideas in Transverse Spin Physics

Dennis Sivers

Portland Physics Institute
Portland, OR 97239

University of Michigan
Ann Arbor, MI 48109

Abstract

Three simple ideas about transverse spin observables are presented for the purpose of stimulating discussion. The manuscript is based on a presentation at the Transversity 2014 Workshop in Torre Chia, Sardinia, Italy on June 9-13, 2014 where approximately sixty experts on transverse spin physics had gathered to share recent results in an atmosphere of sun-drenched intensity.

The title of this talk (chosen by the organizers of the conference) succeeded, for a long interval, in intimidating and baffling the author so that any relevant "ideas" were repelled rather than summoned. Eventually some ideas came but they did not come in any organized fashion. A long history [1-14] of grappling with the challenges of spin observables has enforced a hard-won awareness that progress in this field has, historically, been driven by experimental data.[15] Most of the ideas embodied in the theoretical tools now used to understand these data have emerged sporadically, in awkward stages, as theorists struggled to catch up with well-established experimental results. At the time of this conference, however, a precarious balance between experiment and theory seems to have been reached. The significant progress in the understanding of the factorization and evolution of transverse-momentum-dependent distributions and fragmentation functions (TMD's) that has come to fruition in the past few years, [16,17] has proved wonderfully useful. In consequence of this progress, the TMD formalism now provides solid phenomenological guidance for potential new experimental programs intent on measuring various transverse spin asymmetries as well as other measures of internal hadron structure. The ideas listed below suggest the need for a combination of new experimental data and new approaches to existing data.

*Idea #1. Transverse spin observables can provide useful phenomenological aids that can lead to a more complete understanding of confinement and chiral symmetry breaking in QCD.*
As shown by Kane, Pumplin and Repko[18] (KPR) all transverse single-spin asymmetries in perturbative QCD vanish as $m_q \to 0$. These observables can be seen to require nonperturbative spin-orbit correlations. Conveniently, color confinement and chiral symmetry breaking in the nonperturbative sector of QCD both necessarily produce significant spin-orbit correlations in hadronic systems that can lead to large single-spin asymmetries. Color confinement implies that hadrons (composite systems of quarks and gluons with net color charge zero) provide the physical basis for the nontrivial states of the complete theory and that the quantum theory has a mass gap. In addition, any confining theory produces helicity non-conserving effects involving the confined constituents. Dynamical symmetry breaking interprets the lightest such hadronic state, the $\pi$ meson, as a "pseudo-Goldstone-boson" of the broken chiral symmetry with a mass given by [19]

$$f_\pi m_\pi^2 = -\langle \bar\psi_q \psi_q \rangle_c (m_u + m_d) \quad (1.1)$$

and $f_\pi$ is the pion decay constant. The effective interactions of virtual pions with spin-$\frac{1}{2}$ constituents involve a $\gamma_5$ factor which mixes helicities. The combination of the two nonperturbative dynamical mechanisms necessarily leads to spin-orbit correlations involving the quarks and gluons inside a hadron or within a QCD jet. A quantifiable phenomenological description of these spin-orbit effects in terms of quarks and gluons can justifiably be called "orbital chromodynamics". [7-11] Thus, the pion tornado produced by the transitions,

$$U \uparrow \Rightarrow d \downarrow (\bar d \downarrow u \uparrow)_\pi L = +1, etc. \quad (1.2)$$

found in the Georgi-Manohar [20] chiral-quark model provides a good example of the nonperturbative aspects of orbital chromodynamics. These transitions can be viewed in the context of dynamical origin of constituent quark mass described by the Schwinger-Dyson equations [21] or lattice simulations [22]. The pion tornado described by (1.2) releases this energy (approximately one third of the proton mass) into a virtual $L = 1$ state. The pion tornado produces flavor isospin-dependent dynamics so that, for example

$$|L_u - L_d| \Box |L_u + L_d| \quad (1.3)$$

In the simplest version of the Georgi-Manohar model, $(\vec L_u - \vec L_d) \cdot \hat\sigma_p \geq 0$. The result (1.3) cannot be generated in perturbation theory alone since gluons carry no isospin. The extraction of information about these nonperturbative mechanisms from transverse spin asymmetries necessarily involves phenomenological models but that does not mean that future experiments should be limited by the limited content of existing phenomenology. In particular, it seems a good idea to study transverse spin asymmetries for hadrons produced in the target fragmentation region as well as asymmetries associated with the quark or gluon jets. Also, particle production asymmetries associated with the spin direction of $\Lambda_s^0 \uparrow$ as well as those associated with $\Lambda_c^+ \uparrow$ and $\Lambda_b^0 \uparrow$ based on the tools introduced in ref. 11 can provide significant new insight into the color-rearrangement dynamics leading to hadrons in the central region of QCD jets.

*Idea #2. The spin-directed momentum transfers $\langle \delta k_{TN}(x, \mu^2) \rangle$ found in the formulation of $A_\tau$-odd distribution functions and $\langle \delta p_{TN}(z, \mu^2) \rangle$ for $A_\tau$-odd fragmentation functions provide a way of studying quantum effects associated with the application of one unit of $\hbar$ to different hadronic systems.*

Transverse single-spin asymmetries measure the effect on a system of changing the direction one spin. For asymmetries involving nucleons or quarks, this involves one unit of $\hbar$. The average spin-directed momentum shift designated $\langle \delta k_{TN}(x,\mu^2) \rangle$ (where $k_{TN} = \vec{k}_T \cdot (\hat{s} \times \hat{p})$) found in the transverse-spin asymmetry produced in a hard-scattering process by the $A_\tau$-odd dynamics ($A_\tau$ is called naïve time reversal0 encoded in an $A_\tau$-odd transverse momentum dependent distribution function $\Delta^N G(x, k_{TN}, \mu^2)$. The spin-directed momentum transfer provides a rigorous definition of the P-even $A_\tau$-odd single-spin asymmetry $A_N(x, k_{TN})$ produced by the nonlocal spin-orbit dynamics,

$$\frac{1}{2}\langle \delta k_{TN} \rangle = \frac{-A_N(k_{TN})G^+(k_{TN})}{\partial G^+ / \partial k_{TN}} \quad (1.4)$$

where $G^+(x, k_{TN}, \mu^2)$ is the $A_\tau = +$ spin-averaged distribution integrated over the $k_{TS}$ transverse momentum variable. It is possible to similarly define the transverse momentum shift produced by an $A_\tau$-odd fragmentation function, $\langle \delta p_{TN}(z, \mu^2) \rangle$ by the expression

$$\frac{1}{2}\langle \delta p_{TN} \rangle = \frac{-A_N(k_{TN})D^+(p_{TN})}{\partial D^+ / \partial p_{TN}} \quad (1.5)$$

This approach to spin-directed momentum transfers allows the transition from a TMD formulation of hard-scattering dynamics into an expression for a higher-twist operator with collinear factorization in the hard-scattering process using a simple integration by parts on integral over $k_{TN}$ or $p_{TN}$ in the hard-scattering expression to move the derivatives in (1.4) or (1.5) from the TMDs to the hard-scattering cross section. The concept of spin-directed momentum transfer unifies SSA dynamics in different kinematic regions and also unifies the TMD formulation with the twist expansion.

*Idea #3. It is important to pay attention to the requirement that spin-directed momenta. $\langle \delta k_{TN}(x,\mu^2) \rangle$ generated in hadron-hadron collisions have a significantly different origin from those generated in lepton-hadron collisions.. The very successful RHIC spin program therefore should be enhanced and supplemented by spin physics at other hadronic facilities.*

Unlike the transverse shifts $\langle \delta p_{TN} \rangle$ produced by $A_\tau$-odd fragmentation functions, which are process-independent but rank-dependent, the spin-directed momentum shifts $\langle \delta k_{TN} \rangle$ are strongly process dependent. The well-known result of Collins conjugation [23] can be written in the form

$$\langle \delta k_{TN} \rangle_{SIDIS} = -\langle \delta k_{TN} \rangle_{DY} \quad (1.6)$$

In hadron-hadron collisions there are multiple hard processes involving the exchange of a hard gluon. In contrast to the processes of SIDIS or DY these color exchanges liberate the orbiting quark and the spin asymmetries are produced by the front-back asymmetry of the orbit as described in ref. [7]. Prospective new hadron facilities for transverse-spin physics include the proposal for polarized protons at the Fermilab main injector [24], the NICA spin-physics program [25] described at this conference by Oleg Teryaev and the proposal for a polarized target in AFTER, a fixed target program using crystal extracted beams at the LHC [26]. In studying single-spin asymmetries in hadron-hadron collisions it is particularly interesting to use two-hadron correlations in the final state in order to separate distribution contributions from fragmentation effects as discussed in Ref. [14].

This discussion has benefited from discussions with John Collins, Oleg Teryaev and Christine Aidala.